\documentclass[%
11pt,
 nofootinbib,
 amsmath,amssymb,
 aps,
]{revtex4-1}

\usepackage{graphicx}% Include figure files
\usepackage{dcolumn}% Align table columns on decimal point
\usepackage{bm}% bold math
\usepackage{bbold}
\usepackage{subfigure}
\usepackage{booktabs} % for much better looking tables
\usepackage{array} % for better arrays (eg matrices) in maths
\usepackage{adjustbox}

\usepackage{color}

\begin{document}

\title{Analysis and forecast of COVID-19 spreading in China, Italy and France}

%\preprint{APS/123-QED}

\author{Duccio Fanelli$^{1}$, Francesco Piazza$^{2,3}$ \vspace*{.25cm}}
\affiliation{$^1$Dipartimento di Fisica e Astronomia, Universit\`{a}  di Firenze, INFN and CSDC, 
                 Via Sansone 1, 50019 Sesto Fiorentino, Firenze, Italy}
\email{Duccio.Fanelli@unifi.it}                 
\affiliation{$^2$Centre de Biophysique Moléculaire (CBM), CNRS-UPR 4301, Rue C. Sadron, 45071 Orléans, France}
\affiliation{$^3$ Université d'Orléans, Château de la Source, 45071 Orléans Cedex, France}
\email{Francesco.Piazza@cnrs-orleans.fr}

\begin{abstract}
\noindent In this  note we analyze the temporal dynamics of the 
coronavirus disease 2019 outbreak in China, Italy and France in the time window
$22/01-11/03/2020$.
A first analysis of simple day-lag maps  points to some universality 
in the epidemic spreading, suggesting that simple mean-field models can
be meaningfully used to gather a quantitative picture of 
the epidemic spreading, and notably the height and time of the peak of confirmed 
infected individuals.\\
\indent The analysis of the same data within a simple susceptible-infected-recovered-deaths model
indicates that the kinetic parameter that describes the rate of recovery seems to be   
the same, irrespective of the country, while the infection and death rates appear to be more variable.
The model places the peak in Italy around March 21$^{\rm st}$ 2020, with a
maximum number of  infected individuals of about 15,000 and a 
number of deaths at the end of the epidemics of about 9,300, consistent with
figures typical of seasonal flu epidemics. Since the confirmed cases 
are believed to be between 10 and 20 \% of the real number of individuals who eventually get infected, the apparent 
mortality rate of COVID-19 falls between 3 \% and 7 \% in Italy, while it appears substantially 
lower, between 1 \% and 3 \% in China.  \\
\indent Based on our calculations, we estimate that $2000$ ventilation units 
should represent a fair figure for the peak requirement
to be considered  by  health authorities in Italy for their strategic planning.
\indent Finally, a simulation of the effects of drastic containment measures on the outbreak in Italy
indicates that a reduction of the infection rate 
indeed causes a quench of the epidemic peak. However, it is also seen that 
the infection rate needs to be cut down drastically and quickly to observe an appreciable decrease 
of the epidemic peak and mortality rate. 
This appears only possible through a concerted and disciplined, albeit painful, effort 
of the population as a whole.

%
% QUI C'E' DA CAPIRE MEGLIO SE IL PICCO VIENE SEMORE 1.5 MESI O QUELLO CHE E;
% DOPO COSA? DOPO CHE SI E' COMINCIATO A FARE I TAMPONI (i.e. senso di "confirmed" )
% 
\end{abstract}

\pacs{Valid PACS appear here}% PACS, the Physics and Astronomy
\maketitle

%%%%%%%%%%%%%%%%%%%%%%%%%%%%%%%%%%%%%%%%%%%%%%%%%%%%%%%%%%%%%%%%%%%%%%%%%%%%%%%%%%%%%%%%%%%%%%%%%%%%%%%%%%%%%%%
\section{Introduction}

\noindent In December 2019 coronavirus disease 2019 (COVID-19) emerged in Wuhan, China.
Despite the drastic, large-scale containment measures promptly implemented by the Chinese
government,  in a matter of a few weeks the disease had spread well outside China,
reaching countries in all parts of the globe. Among the countries hit 
by the epidemics, Italy found itself grappling with the worst outbreak after the original one,
generating considerable turmoil among the population.
The exponential increase in people who tested positive to COVID-19 (supposedly together with the 
sudden increase in the testing rate itself), finally  
prompted the Italian government to issue on March 8$^{\rm th}$ 2020 a dramatic decree
ordering the lockdown of the entire country. \\
\indent In this technical note, we report the results of a comparative assessment 
of the evolution of COVID-19 outbreak in mainland China, Italy and France.
Besides shedding light on the dynamics of the epidemic spreading, 
the practical intent of our analysis is to provide officials with realistic 
estimates for the time and magnitude 
of the epidemic peak, i.e. the maximum number of infected individuals,
as well as gauge the effects of drastic containment measures
based on simple quantitative models.  
Data were gathered from the github repository associated with the interactive dashboard 
hosted by the Center for Systems Science and Engineering (CSSE) at Johns Hopkins University, 
Baltimore, USA~\cite{Dong:2020aa}. The data analyzed in this study correspond to the
period that stretches between  January 22$^{\rm nd}$ 2020 and March 11$^{\rm th }$ 2020, included.\\
%

%%%%%%%%%%%%%%%%%%%%%%%%%%%%%%%%%%%%%%%%%%%%%%%%%%%%%%%%%%%%%%%%%%%%%%%%%%%%%%%%%%%%%%%%%%%%%%%%%%%%%%%%
\section{Preliminary insight from recurrence plots}

\begin{figure}[!t]
\centering
\includegraphics [width=10truecm]{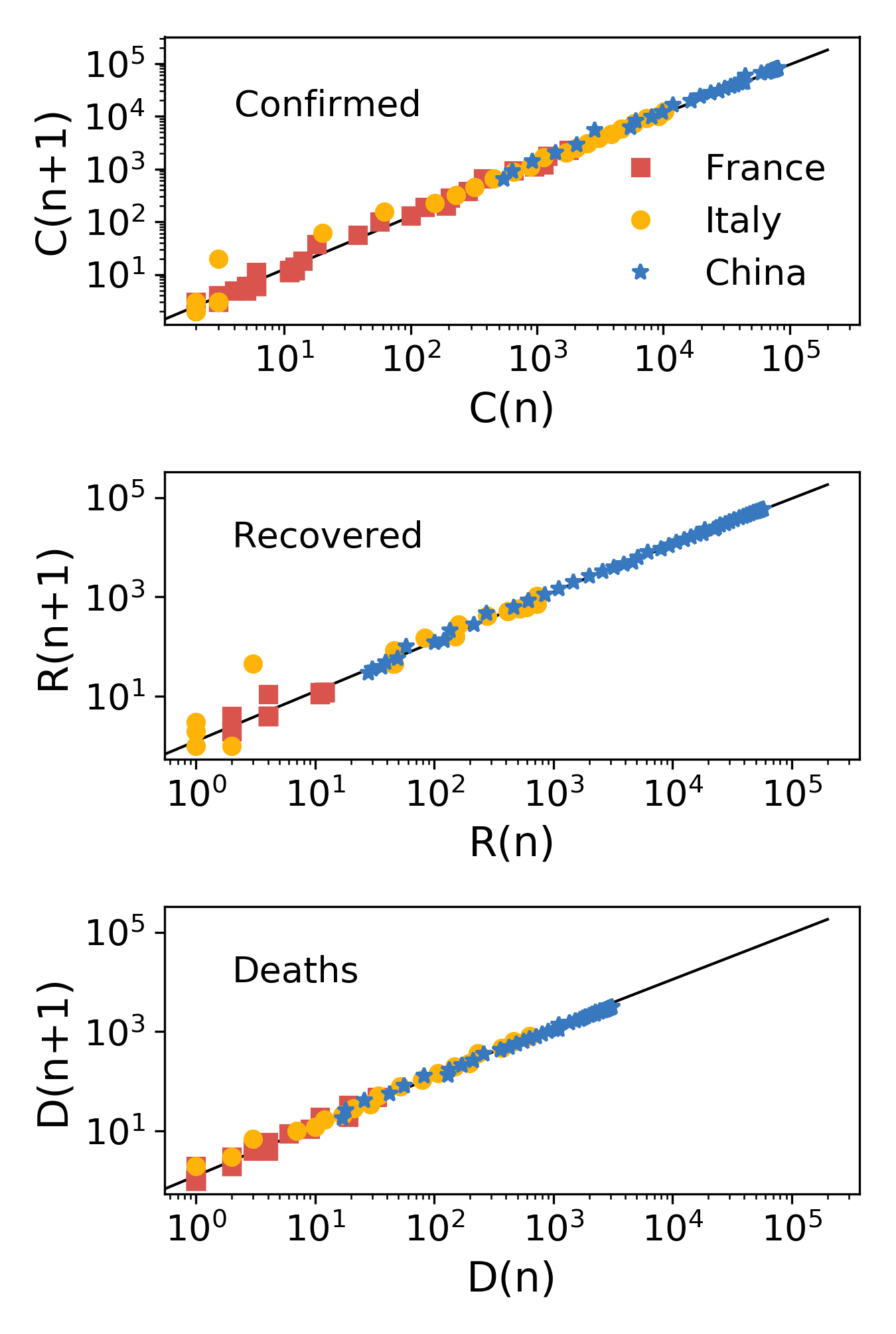}
\caption{Recurrence plots for the three populations for which data ara publicly available (symbols)
for the outbreaks in China, Italy and France and best fit with a power law of the 
kind~\eqref{e:pl} (solid lines). All data appear to follow the exact same trend on average (see text).}
\label{f:map}
\end{figure}

\noindent A first simple analysis that can be attempted to get some insight into 
the outbreak dynamics is to build iterative time-lag maps. The idea is to investigate the 
relation between some population  at time (day) $n+k$ and the same population at day $n$,
corresponding to a time lag of $k$ days. Of course, the simplest case of all is
to build  day-by-day maps ($k=1)$. 
We built three such maps, associated with the population of cumulative confirmed infected 
people ($C$), recovered people ($R$) and total reported deaths ($D$) for the three countries 
considered. We note that $I = C-(R+D)$ is the total number of infected individuals,
i.e. without taking into accounts recoveries and deaths.
Fig.~\ref{f:map}
shows that in all cases the data follow the same power law of the kind  
\begin{equation}
\label{e:pl}
P_{n+1}=\alpha P_{n}^{\,\beta}
\end{equation}
where $\alpha = 2.173$ and  $\beta = 0.928$ and $P=(C,R,D)$. 
This observation suggests that there is some universality in the epidemic spreading 
within each country. As a consequence, simple models of the mean-field kind can
be adopted to gather a meaningful and quantitative picture of the epidemic spreading 
in time, to a large extent irrespective of the specific country of interest.
In the second part of this note, we provide a concrete example of such an analysis
for  two of the three countries considered here.\\
\indent It should be noted that the predicted time evolution of the three populations can be 
computed analytically from the iterative 
map~\eqref{e:pl} (see appendix). More precisely, one has 
\begin{equation}
\label{e:Pn}
P_n=\alpha^{(1-\beta^n)/(1-\beta)} P_0^{\beta^n} 
\end{equation}
where $P=(C,R,D)$.
Reassuringly, we find $\beta<1$, which means that the sequence~\eqref{e:Pn} converges to a 
plateau, which is the (stable) fixed point of the function $F(x) = \alpha x^\beta$. 
Hence, for any value of $P_0>0$, one has
\begin{equation}
\label{e:fp}
\lim_{n\to\infty} P_n = \alpha^{1/(1-\beta)} 
\end{equation}
It should be observed that the three populations $C$, $R$ and $D$ are expected to 
level off at three different values. With respect to Eq.~\eqref{e:fp}, this simply means 
that one should regard $\beta = 0.928$ as an average figure. In fact, each 
population will be characterized by slightly different value of  $\beta$, which will
yield considerably different plateaus, since they are all close to the singularity at $\beta =1$
(see again Eq.~\eqref{e:fp}).
%
%%%%%%%%%%%%%%%%%%%%%%%%%%%%%%%%%%%%%%%%%%%%%%%%%%%%%%%%%%%%%%%%%%%%%%%%%%%%%%%%%%%%%%%%%%%%%%%%%%%%%%%%%%%%%
%%%%%%%%%%%%%%%%%%%%%%%%%%%%%%%%%%%%%%%%%%%%%%%%%%%%%%%%%%%%%%%%%%%%%%%%%%%%%%%%%%%%%%%%%%%%%%%%%%%%%%%%%%%%%%
%
The prediction~\eqref{e:fp} should not be regarded as the true asymptotic value
to be expected at the end of the outbreak for either populations. Rather, it should be 
regarded as an estimate  of the total population initially within the ensemble of people 
who will eventually get infected.
In fact, the elements of the ensemble
$(C,R,D)$ are not independent, as people get infected, recover and die 
as time goes by, thus effectively transferring elements from one population 
to another. We will show in the next section how this can be accounted for 
within a simple {\em kinetic} scheme, where eventually 
such interactions will cause the population of infected individuals $I$ to die out and the $R$ and $D$
populations to reach two separate plateaus as observed.  
Furthermore, it should be 
noticed that the data plotted in Fig.~\ref{f:map} start from the 
first pair of successive values $(P_{n+1},P_n)$ encountered in the data sheets 
with $P_{n+1},P_n>0$, consistent with the fact that $P=0$ is also a (trivial)
fixed point of the map~\eqref{e:pl}. \\
%
% 
%
%%%%%%%%%%%%%%%%%%%%%%%%%%%%%%%%%%%%%%%%%%%%%%%%%%%%%%%%%%%%%%%%%%%%%%%%%%%%%%%%%%%%%%%%%%%%%%%%%%%%%%%%%%%%%%%%%%%%%%%%%%%%%%%
\begin{figure}[!t]
\centering
\includegraphics [width=12truecm]{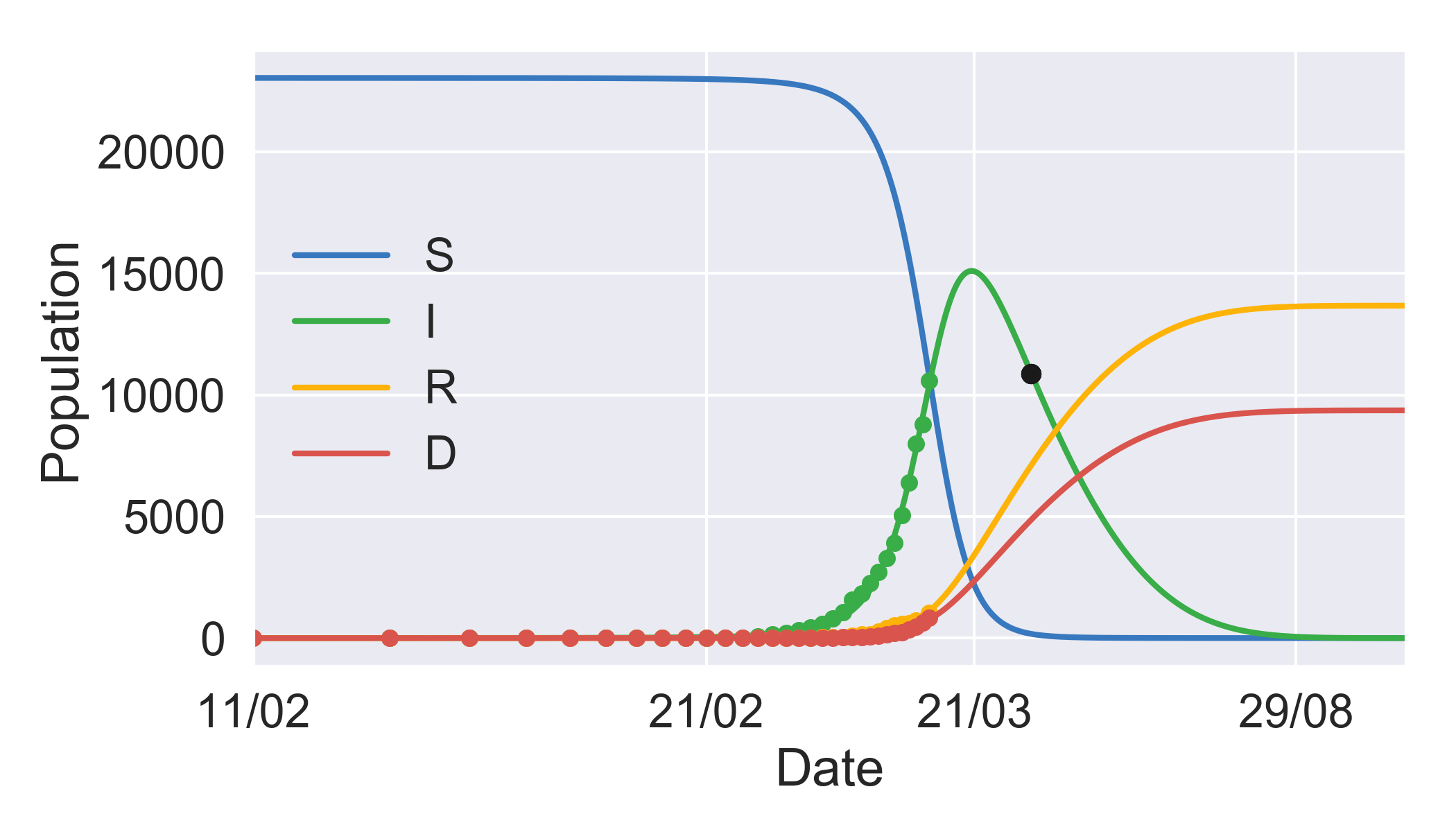}
\includegraphics [width=12truecm]{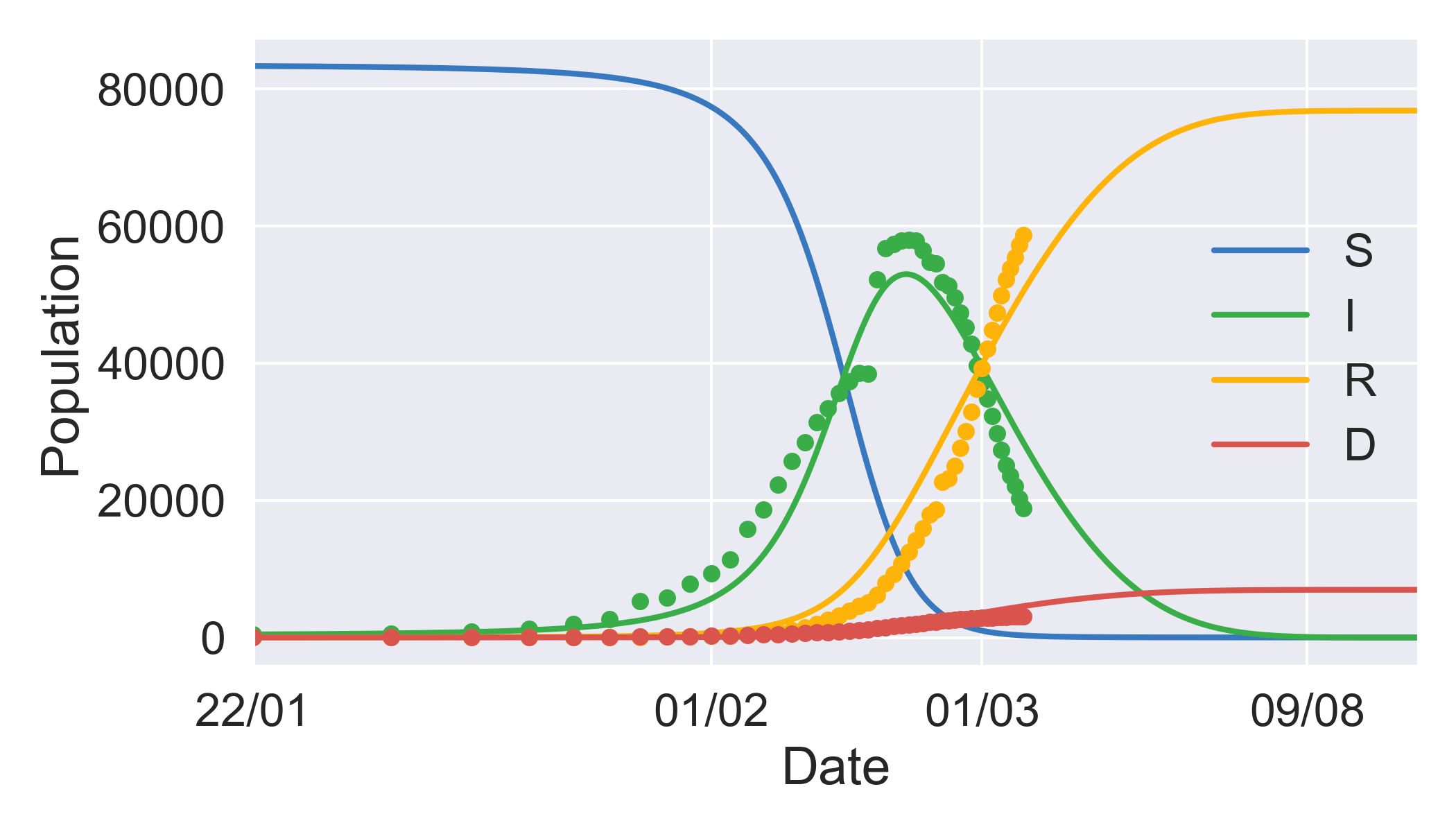}
\caption{Predicted evolution of the COVID-19 outbreak in Italy (top) and China (bottom).
Symbols represent the official data retrieved from the CSSE repository~\cite{Dong:2020aa}.
Solid lines are the predicted trends based on the fits of the SIRD model, Eqs~\eqref{e:SIRD}, to the data.
The black circle in the top graph marks the predicted number of confirmed infected individuals 
at the announced end of the imposed lockdown on the Italian territory, April 3$^{\rm rd}$ 2020.}
\label{f:SCRD}
\end{figure}
%%%%%%%%%%%%%%%%%%%%%%%%%%%%%%%%%%%%%%%%%%%%%%%%%%%%%%%%%%%%%%%%%%%%%%%%%%%%%%%%%%%%%%%%%%%%%%%%%%%%%%%%%%%%%%%%%%%%%%%%%%%%%%%
%
% 

%========================================================================================================
\section{Mean-field kinetics of the epidemic spreading: exponential growth, peak and decay}

\noindent As more people get infected, more people also recover or, unfortunately, die. 
Within the simplest model of the  evolution of   
an epidemic outbreak, people can be divided into different classes (species).
In the  susceptible (S),  infected (I), recovered (R), dead (D) 
scheme (SIRD), any individual in the fraction of the overall population that will
eventually get sick belongs  to one of the aforementioned classes. Let $S_0$ be 
the size of the initial population of susceptible people. 
The mean-field~\footnote{In a mean-field approach such as this one, spatial effects 
are neglected, while the populations are considered as averaged over the 
whole geographical scene of the epidemics outbreak. This is  
much like the concept of average concentrations of reactants 
when the assumption of a well-stirred chemical reactor is made
in chemical kinetics.} kinetics of the SIRD epidemic evolution is described by the 
following system of differential equations 
\begin{eqnarray}
 \label{e:SIRD}
  \frac{dS}{dt} &=& -r S I \nonumber\\
  \frac{dI}{dt} &=&  r S I - (a+d) I \nonumber\\
  \frac{dR}{dt} &=&  a  I \nonumber\\
  \frac{dD}{dt} &=&  d  I \nonumber\\
 \end{eqnarray}
with initial condition $[S(t_0),I(t_0),R(t_0),D(t_0)] = [S_0,I_0,R_0,D_0]$ for some
initial time $t_0$. The parameter $r$ is the 
infection rate, i.e. the probability per unit time that a susceptible individual contract
the disease when entering in contact with an infected person.
The parameters $a$ and $d$ denote, respectively, the recovery and death rates.
Although the SIRD model is rather crude, the kind of universality emerging from the analysis 
reported in the previous section for the evolution of non-interacting populations
suggests that such scheme has good chances to capture at least the gross 
features of the full time course of the outbreak. \\
\indent Fig.~\ref{f:SCRD} illustrates the results of fitting the (numerical)
solution of Eqs.~\eqref{e:SIRD} simultaneously to the 
data for the three populations reported in the CSSE sheets, i.e. $I(t), R(t)$  and $D(t)$,
for the outbreaks in China and Italy. We found that the data reported for the outbreak 
in France are still too preliminary to warrant a meaningful fit of this kind. 
The set of free parameters and the initial conditions used 
were $[r,a,d,S_0]$, $[S_0,I_0,0,0]$ in the case of Italy and
$[r,a,d,S_0,R_0,D_0]$, $[S_0,I_0,D_0,R_0]$ in the case of China, respectively. 
In the former case, due to the prolonged initial stretch of stagnancy (presumably due to the initial 
low testing rate), we  chose $t_0 = 20$ days after day one (22/01/2020) and fixed 
the populations at the corresponding reported values $I_0 = 3, R_0=D_0=0$. In the case of China,
we set $t_0$ to day one, as the reported initial populations bear evidence
of an outbreak that is already well {\em en route}. However, we found that the initial 
values reported for all the populations, but notably the  infected individuals,
appear underestimated.  This is consistent with the abrupt, visible increase appearing 
around mid-February, when  Chinese authorities changed the testing protocol~\cite{Dong:2020aa,WHO}.
Consequently, we let the initial values $I_0,S_0,D_0$ float as well during the fits. 
Interestingly, we found that identical fits could be obtained by fixing the initial values of the 
populations  at the reported values and allowing for a (negative) lag time $\tau$, signifying 
a shift in the past of the {\em true} time origin of the epidemics.  In this 
case, we obtained  $\tau = 30$ days, consistent with the presumed outset of the outbreak.\\
%
%%%%%%%%%%%%%%%%%%%%%%%%%%%%%%%%%%%%%%%%%%%%%%%%%%%%%%%%%%%%%%%%%%%%%%%%%%%%%%%%%%%%%%%%%%%%%%%%%%%%%%%%%%
%
% Requires the booktabs if the memoir class is not being used
\begin{table}[t!]
   \centering
   \begin{footnotesize}% or footnotesize, scriptsize, tiny, etc.
   \begin{tabular}{@{} lccccccc @{}} % Column formatting, @{} suppresses leading/trailing space
      \hline
      {\bf Country} & $r$ [days$^{-1}$]  & $\ \ $ &
                      $a$ [days$^{-1}$]  & $\ \ $ &
                      $d$ [days$^{-1}$]  & $\ \ $ & 
                      $S_0$  \\
      \hline\hline
      {\bf Italy}  & $1.460\times10^{-5} \pm 5\times10^{-8}$  & $\ \ $  &
                $2.13\times10^{-2}  \pm 2\times10^{-4}$  & $\ \ $ &
                $1.45\times10^{-2}  \pm 2\times10^{-4}$  & $\ \ $ &
                $2.303\times10^{4}   \pm 8\times10^{1}$   \\
      {\bf China}   & $3.95\times10^{-6}  \pm 4\times10^{-8}$  & $\ \ $ &
                $3.53\times10^{-2}  \pm 1\times10^{-4}$  & $\ \ $ &
                $3.1\times10^{-3}  \pm 2\times10^{-4}$   & $\ \ $ &
                $8.33\times10^{4}  \pm 2\times10^{2}$ \\
      {\bf China}$^\ast$ & $3.33\times10^{-6}  \pm 2\times10^{-8}$  & $\ \ $ &
                $1.80\times10^{-2}  \pm 2\times10^{-4}$  & $\ \ $ &
                $3.0\times10^{-3}  \pm 2\times10^{-4}$   & $\ \ $ &
                $7.92\times10^{4}  \pm 4\times10^{2}$ \\    
      \hline
   \end{tabular}
   \end{footnotesize}
   \caption{Table of average values of the best-fit parameters and associated standard deviations
   computed from 30 independent runs of the stochastic differential evolution algorithm~\cite{Storn:1997aa}, as
   implemented in the \texttt{Python-Scipy} package. The line marked with an asterisk refers to a fit 
   limited to the data up to February 19$^{\rm th}$ 2020.}
   \label{t:pars}
\end{table}
%
%
%%%%%%%%%%%%%%%%%%%%%%%%%%%%%%%%%%%%%%%%%%%%%%%%%%%%%%%%%%%%%%%%%%%%%%%%%%%%%%%%%%%%%%%%%%%%%%%%%%%%%%%%%%%%%%%%%%%%%%%%%%%%%%%
\begin{figure}[!t]
\centering
\includegraphics[width=12truecm]{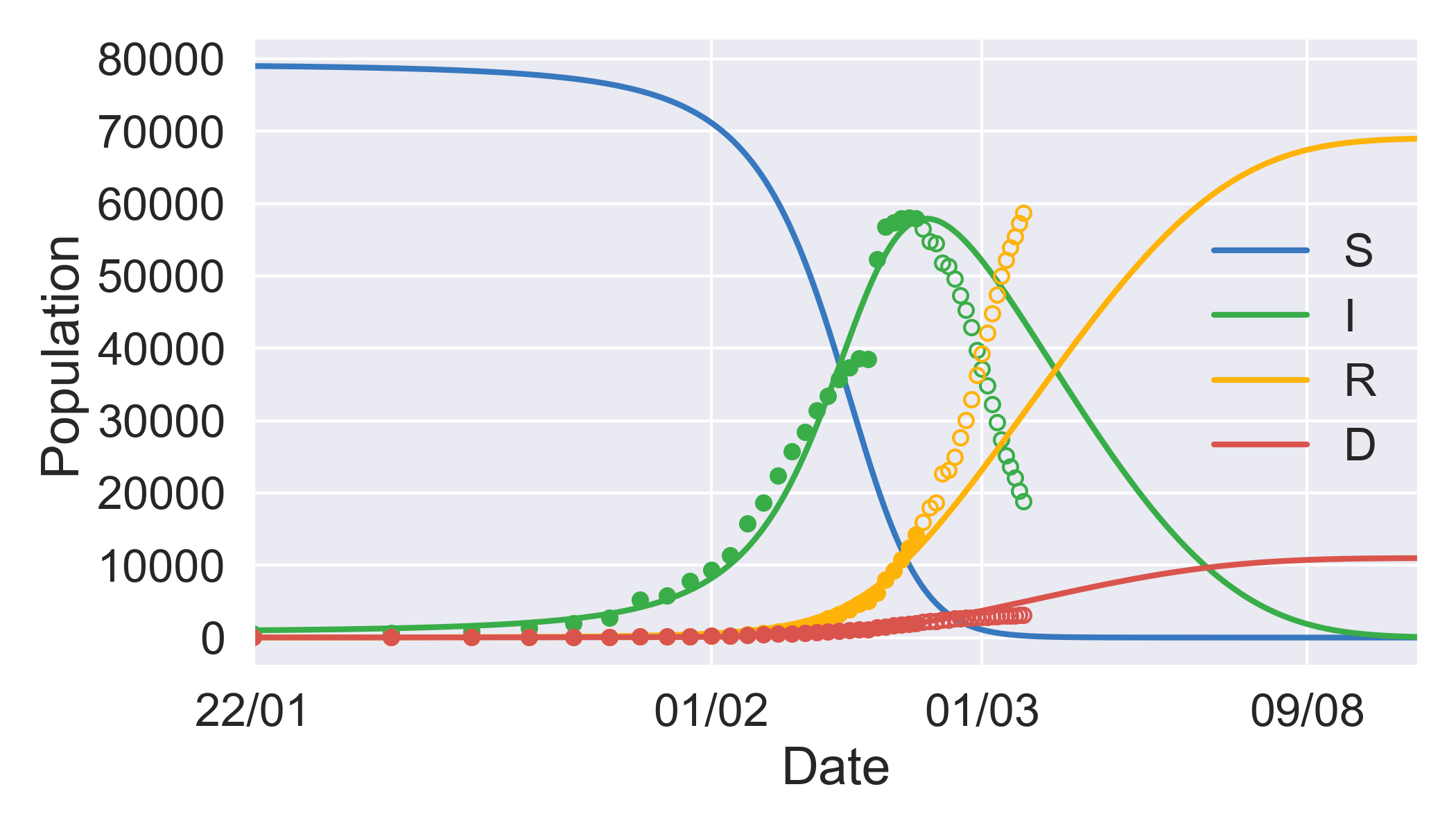}
\caption{Predicted evolution of the COVID-19 outbreak in  China obtained by fitting 
the data up to to February 19$^{\rm th}$ 2020. The fitted data are shown as filled circles 
(see also Table~\ref{t:pars}). A very similar prediction
is obtained by restricting the fit up to  February 15$^{\rm th}$ 2020,
where the peak had not been reached yet (data not shown).}
\label{f:fitfinopicco}
\end{figure}
%%%%%%%%%%%%%%%%%%%%%%%%%%%%%%%%%%%%%%%%%%%%%%%%%%%%%%%%%%%%%%%%%%%%%%%%%%%%%%%%%%%%%%%%%%%%%%%%%%%%%%%%%%%%%%%%%%%%%%%%%%%%%%%
%
\indent The best-fit values of the floating  parameters are listed in Table~\ref{t:pars}.
We find that the recovery rate does not seem to depend on the country, while 
the infection and death rate show a more marked variability. This is likely to be connected 
with many culture-related habits and to the presumed diversity in underlying health conditions of 
the more vulnerable that are expected to influence these parameters.
It should also be noted that this discrepancy might eventually get reduced when more data on the 
outbreak in Italy 
will become available. This would also imply an increase of the initial number of susceptible 
people, $S_0$. However,  it turns out that this would entail only a modest shift 
of the epidemic peak forward in time (data not shown here).
The best-fit values of the additional parameters fitted for the  China outbreak were
$I_0 = 430 \pm 20$, $R_0 = 10 \pm 10$, $D_0 = 15 \pm 7$ (full range) and 
$I_0 = 999 \pm 1$, $R_0 = 10 \pm 10$,  $D_0 = 17 \pm 7$ (full range). \\
\indent It can be remarked from Fig.~\ref{f:SCRD} that the global fit of the SIRD model, while predicting 
the observed position of the epidemic peak, it does so 
at the price of a worse interpolation of the initial growth and of the final
decay of the $I$ population. Concurrently, the model fails to follow the observed rapid
recovery and overestimates the number of deaths. This is most likely due to the harsh 
containment measures adopted by the Chinese government in order to curb the spread of the 
disease. A simple way to test this hypothesis is to restrict the fit  to the initial 
growth phase  before the onset of the peak. This is illustrated in 
Fig.~\ref{f:fitfinopicco}. Indeed, it can be appreciated that a model that does not 
include any external curbing action on the infected population
reproduces quite nicely the initial growth phase, places the peak at the correct time, 
but fails to match the swift recovery rate and decline of the infection in the
window where the imposed restrictions are assuredly in action.\\
\indent The analysis of the outbreak in China strongly suggests that the prediction of 
our  nonlinear fitting strategy for the epidemic peak  in Italy is a robust one.
However, most likely these data do not bear any signature yet of the  
harsh, draconian measures  contained in the 
dramatic decree signed by Mr Conte on March 8$^{\rm th}$ 2020.
Equipped with our robust estimates of the kinetic parameters, we are in a good position 
to inquire whether those measures
will impact  substantially on the future evolution of the epidemics. 
To this aim, we consider a modified 
version of the SIRD model, where the  infection rate $r$ is let vary with time.
More precisely, given that the containment measures became law at time $t^\ast$, 
we take
\begin{equation}
\label{e:rt}
r(t) = \begin{cases}
       r_0 & \text{for} \quad t\leq t^\ast\\
       r_0(1-\alpha) e^{-(t-t^\ast)/\Delta t} + \alpha \, r_0 &
       \text{for} \quad t >  t^\ast
      \end{cases} 
\end{equation}
% 
%
%%%%%%%%%%%%%%%%%%%%%%%%%%%%%%%%%%%%%%%%%%%%%%%%%%%%%%%%%%%%%%%%%%%%%%%%%%%%%%%%%%%%%%%%%%%%%%%%%%%%%%%%%%%%%%%%%%%%%%%%%%%%%%%
\begin{figure}[!t]
\centering
\includegraphics [width=12truecm]{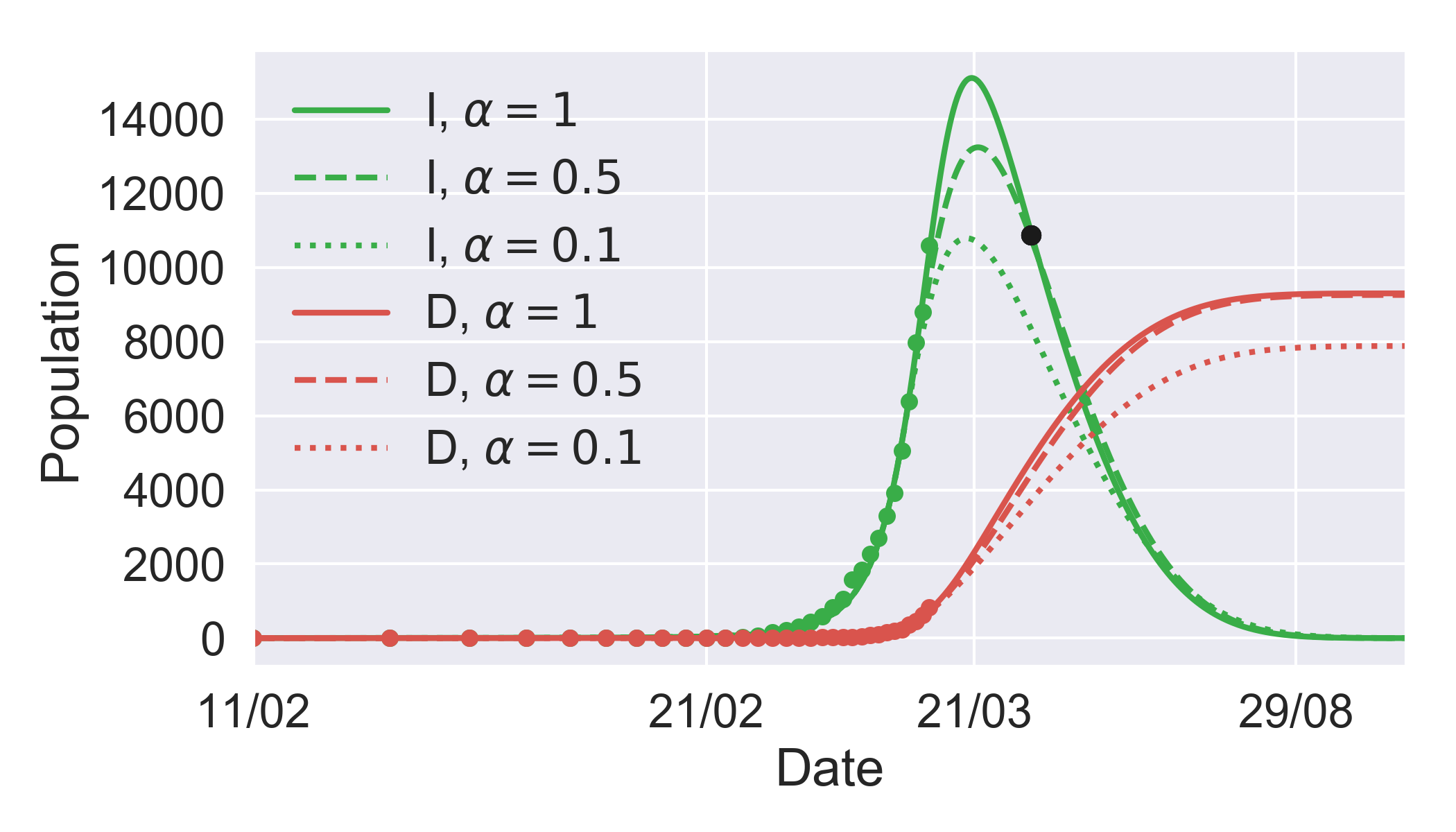}
\includegraphics [width=12truecm]{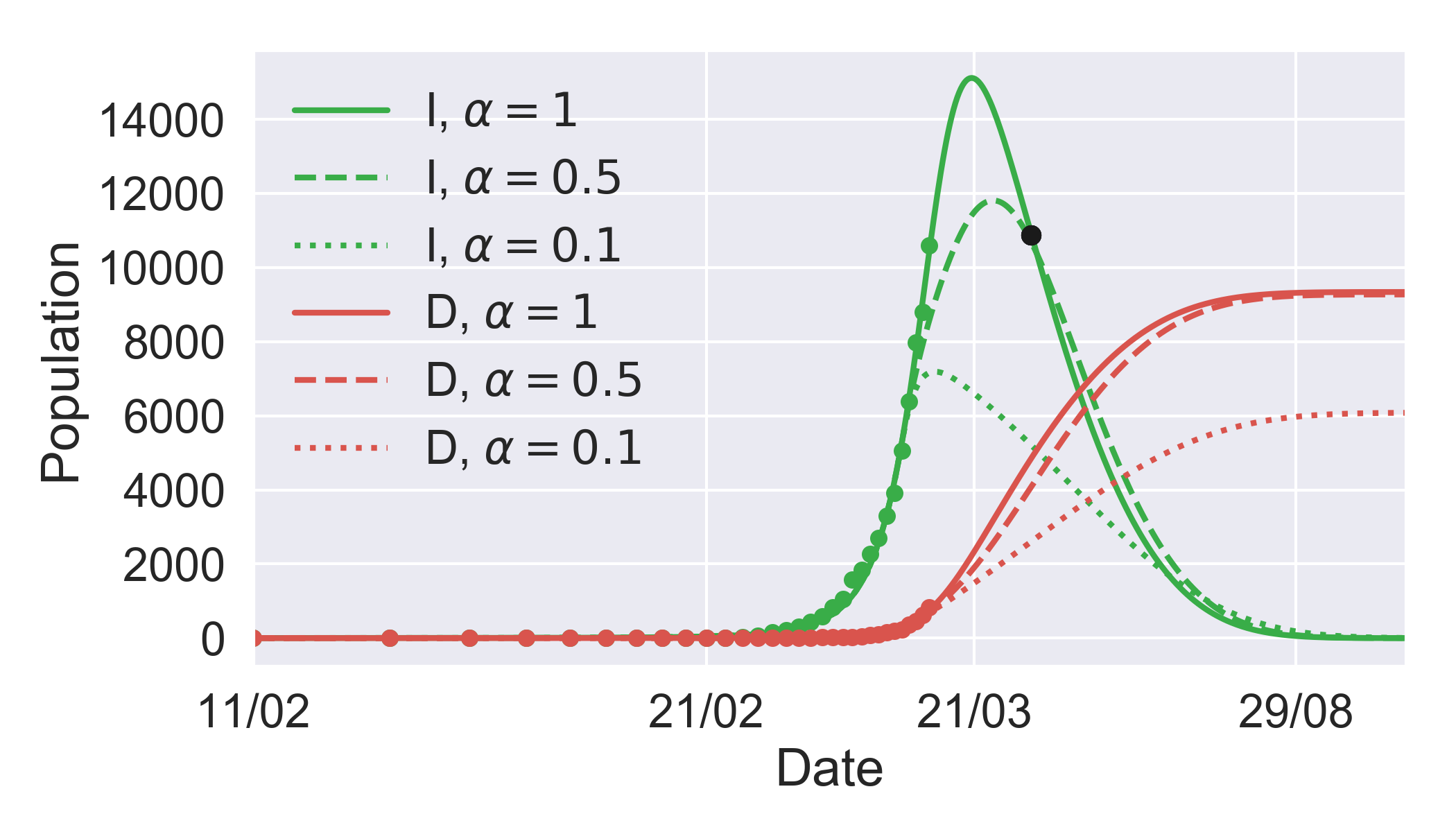}
\caption{Predicted effect of the lockdown measures imposed by the Italian 
government on the whole national territory on March 8$^{\rm th}$ 2020.
The predicted evolution of the confirmed infected population and the number 
of casualties are plotted for different values of the reduction 
of the infection rate achieved thanks to the lockdown, see Eq.~\eqref{e:rt}. 
The black circle  marks the announced end of the imposed lockdown, April 3$^{\rm rd}$ 2020. 
Top graph: $\Delta t = 7$ days. Bottom graph: $\Delta t = 2$ days. }
\label{f:lockdown}
\end{figure}
%%%%%%%%%%%%%%%%%%%%%%%%%%%%%%%%%%%%%%%%%%%%%%%%%%%%%%%%%%%%%%%%%%%%%%%%%%%%%%%%%%%%%%%%%%%%%%%%%%%%%%%%%%%%%%%%%%%%%%%%%%%%%%%
%
where $r_0 = 1.46 \times 10^{-5}$ days$^{-1}$ is the rate estimated from the fit to the data shown 
in Fig.~\ref{f:SCRD}, hence unaffected by the lockdown, and 
$\alpha \in [0,1]$ gauges the asymptotic reduction of the infection rate 
afforded by the containment measures. Fig.~\ref{f:lockdown} shows two 
predictions based on such modified SIRD model, for intermediate (50 \%) and 
large (90 \%) reduction of the infection rate, with $t^\ast$ fixed at the 
date of the signature of the decree and $\Delta t$ = 7 and 2 days, i.e. assuming that 
the effects of the lockdown will be visible on a time of the order of one week or 
a few days.  \\
\indent It can be appreciated that the effect is predicted to be 
the one the government was hoping for. Moreover, it can be seen 
that the quickest the drop in the infection rate brought about by the
containment measures, the more substantial the reduction of the epidemic peak. 
However, it can also be seen that the infection rate should be cut down 
rather drastically for the measures to be effective. 
Overall, the dynamics of the decay of the epidemics after the peak
and the mortality rate seem also little affected by a time decay of the 
infection rate, unless this happens very quickly (in a matter of days) 
and suppressing new infections by at least 90 \%.

%%%%%%%%%%%%%%%%%%%%%%%%%%%%%%%%%%%%%%%%%%%%%%%%%%%%%%%%%%%%%%%%%%%%%%%%%%%%%%%%%%%%%%%%%%%%%%%%%%%%%%%%%
\section{Discussion}

% Time-lag maps ---> universality
\noindent In this report we have analyzed epidemic data made available to the scientific community 
by the Center for Systems Science and Engineering at Johns Hopkins University~\cite{Dong:2020aa}
and referring to the period $22/02/2020 - 11/03/2020$.
Our results seem to suggest that there is a certain universality
in the time evolution of COVID-19. This is demonstrated by time-lag plots of the 
confirmed infected populations of China, Italy and France, which collapse on one 
and the same power law on average. This suggests that a country that becomes the theatre of an epidemics surge 
can be regarded, at least in first approximation, 
as a well-stirred chemical reactor, where different populations 
interact according to mass-action-like rules with little connection to geographical variations. \\
%
% SIRD fits ---> predictions of peak times and cases 
\indent The analysis of the same data within a simple susceptible-infected-recovered-deaths 
(SIRD) model reveals that the recovery rate is the same for Italy and China, while 
infection and death rate appear to be different.
A few observations are in order. Chinese authorities have 
tackled the outbreak by imposing martial law to a large fraction on the population, thus 
presumably cutting down the infection  rate to a large extent. While data on the outbreak 
in China bear the signature of this  measure, the data on the outbreak in Italy clearly do not 
at this stage. Moreover, it can be surmised that many cultural factors could influence the 
infection rate, thus leading to a larger variability from one country to another. 
Analysis of data from more than two countries of course are needed to substantiate this
hypothesis. The death rate probably reflects the average age and underlying health conditions 
of elderly patients, which are also likely to vary markedly depending on culture and lifestyle.\\
\indent As more data will become available for the 
outbreak in France, the same analysis will be attempted on those data too. 
In fact, the outbreak appears to have started later in France than in Italy. However, 
the confirmed cases reported could be biased by a non-stationary testing rate, which 
could have increased substantially after the severity of the outbreak in Italy came under the spotlight. 
This document will be updated regularly during the outbreak, and predictions of the 
peak time and severity in France will be included as soon as the data will make these calculations 
meaningful. \\
\indent The SIRD model places the peak in Italy around March 21$^{\rm st}$ 2020, and predicts a 
maximum number of confirmed infected individuals of about 15,000 at the peak of the outbreak. 
The  number of deaths at the end of the epidemics appear to be about 9,300, which is consistent with
figures typical of seasonal flu epidemics. Taking into account that the confirmed cases 
can be estimated to be between 10 and 20 \% of the real number of infected individuals~\cite{WHO}, 
the apparent  mortality rate of COVID-19 seems to be between 3 \% and 7 \% in Italy, higher 
than seasonal flu, while it appears 
substantially lower in China, that is, between 1 \% and 3 \%. \\
\indent Furthermore, assuming that the fraction of sick people needing intensive care with ventilation 
appears to be about $5-10$ \% of those who contract the disease~\cite{Guan:2020aa}, 
the maximum number of individual ventilation units 
required overall to handle the epidemic peak in Italy, i.e around 15,000 cases, 
can be estimated to be around $1000-1500$. We believe that a more conservative estimate of 
2000 ventilation units as the peak requirement 
represents a fair figure to be handled to the health authorities for their strategic planning.   \\
\indent Finally, based on the kinetic parameters fitted on the data for the outbreak in Italy, 
i.e. up to the day following the painful lockdown of the whole nation enforced on March 8$^{\rm th}$ 2020, 
we have computed the prediction of the SIRD model modified by the highly awaited effects of a 
fading infectivity following the lockdown. While a reduction in the epidemic peak and mortality rate are indeed 
observed,  we predict that such effects will only be visible if the measures cause 
a  quick (matter of days) and drastic (down by at least $80-90$ \%) cutback of the infection rate. 
In Italy and in other countries that will be facing the epidemic surge soon, 
this is quite possibly only achievable  through a cooperative and disciplined effort of the population 
as a whole.\\

\smallskip
\noindent This note is available as an ongoing project on ResearchGate at the following address:\\
\smallskip
\begin{footnotesize}
\noindent\texttt{https://www.researchgate.net/project/Analysis-and-forecast-of-COVID-19-spreading-in-China-and-Europe}
\end{footnotesize}
\noindent The analyses presented in this report will be updated regularly during the course of 
the global outbreak and extended to other countries. The authors hope that this project will 
be of some help to health and political authorities during the difficult moments of this global 
outbreak.

\acknowledgements   \noindent We would like to thank Marco Tarlini for pointing out the 
correct definition of the confirmed infected cases in the CSSE data sheets. 
We are also indebted to the many colleagues who quickly sent us insightful observations
on the pre-print.

%%%%%%%%%%%%%%%%%%%%%%%%%%%%%%%%%%%%%%%%%%%%%%%%%%%%%%%%%%%%%%%%%%%%%%%%%%%%%%%%%%%%%%%%%%%%%%%%%%%%%%%%%%%%%
% APPENDIX

\appendix

\section{Calculation of the explicit form of the iterative map}

\noindent From Eq.~\eqref{e:pl} one can determine the explicit form of the population $P_n$ at time $n$
(P=C,R,D). The steps of the iteration can be worked out explicitly, that is,
\begin{eqnarray*}
P_1 &=&\alpha P_0^{\,\beta} \nonumber \\ 
P_2 &=&\alpha P_1^{\,\beta} = \alpha^{1+\beta} P_0^{\,\beta^2} \nonumber\\
    &\dots& \nonumber\\
P_n &=& \alpha^{1 + \beta + \beta^2 \dots \beta^{n-1}} P_0^{\beta^n}  
\end{eqnarray*}
Recalling that 
\begin{equation}
1 + \beta + \beta^2 \dots + \beta^{n-1 }= \frac{1-\beta^n}{1-\beta}
\end{equation}
one eventually gets Eq.~\eqref{e:Pn}.

%%%%%%%%%%%%%%%%%%%%%%%%%%%%%%%%%%%%%%%%%%%%%%%%%%%%%%%%%%%%%%%%%%%%%%%%%%%%%%%%%%%%%%%%%%%%%%%%%%%%%%%%%%%%
% BIBLIOGRAPHY

\bibliography{corona}

\end{document}